%% file: ms.tex
\documentclass[12pt,preprint]{aastex}
\usepackage{lscape,graphicx,fullpage}

\usepackage{epsfig}

\include {new-noteset}

\def\plotsix#1#2#3#4#5#6{\centering \leavevmode
    \epsfxsize=.160\textwidth \epsfbox{#1}
    \hfil \epsfxsize=.160\textwidth \epsfbox{#2}
    \hfil \epsfxsize=.160\textwidth \epsfbox{#3}
    \hfil \epsfxsize=.160\textwidth \epsfbox{#4}
    \hfil \epsfxsize=.160\textwidth \epsfbox{#5}
    \hfil \epsfxsize=.160\textwidth \epsfbox{#6}}

\title{On the nature and correction of the spurious S-wise spiral galaxy winding bias in Galaxy Zoo 1}
\author{Wayne B. Hayes, Darren Davis, Pedro Silva}
\affil{Computer Science Department University of California, Irvine\\Irvine, California 92697-3435}
\email{whayes@uci.edu}

\begin{abstract}
The Galaxy Zoo 1 catalog displays a bias towards the S-wise winding direction in spiral galaxies which has yet to be explained. The lack of an explanation confounds our attempts to verify the Cosmological Principle, and has spurred some debate as to whether a bias exists in the real universe. The bias manifests not only in the obvious case of trying to decide if the universe as a whole has a winding bias, but also in the more insidious case of selecting which galaxies to include in a winding direction survey. While the former bias has been accounted for in a previous image-mirroring study, the latter has not. Furthermore, the bias has never been {\em corrected} in the GZ1 catalog, as only a small sample of the GZ1 catalog was re-examined during the mirror study. We show that the existing bias is a human {\em selection} effect rather than a human chirality bias. In effect, the excess S-wise votes are spuriously ``stolen'' from the elliptical and edge-on-disk categories, not the Z-wise category. Thus, when selecting a set of spiral galaxies by imposing a threshold $T$ so that $\max(P_S,P_Z) > T$ or $P_S+P_Z>T$, we spuriously select more S-wise than Z-wise galaxies. We show that when a provably unbiased machine selects which galaxies are spirals independent of their chirality, the S-wise surplus vanishes, even if humans are still used to determine the chirality. Thus, when viewed across the entire GZ1 sample (and by implication, the Sloan catalog), the winding direction of arms in spiral galaxies as viewed from Earth is consistent with the flip of a fair coin.
\end{abstract}

\begin{document}
\section{Introduction}
\label{sec:intro}

The {\it Cosmological Principle} is the assumption that at large scales
the universe is homogeneous and isotropic.  Homogeneity says that there
is no special location in the Universe, and in particular that the Earth
occupies no special location.  Isotropy means that there is
no preferred direction in the universe; for spiral galaxies, this means that the distribution 
of their spin axes should be spread uniformly at random on the celestial sphere.
The two assumptions together imply that, as seen from the Earth, the
distribution of observed arm winding directions of $N$ spiral galaxies
should be statistically consistent with $N$ flips of a fair coin.

The Galaxy Zoo 1 (hereafter GZ1) project \citep{Lintott2008,Lintott2010} was a website where humans
were presented with random galaxy images from the Sloan Digital Sky Survey
\citep{SDSS}. With each galaxy image they were given a choice of 6 ``cartoon'' galaxies
and asked which cartoon most resembled the real
galaxy.  The GZ1 sample has almost 900,000 galaxies.
After using SpArcFiRe \citep{DavisHayes2014} to perform an ellipse fit of
all GZ1 images, we concentrate on
a subsample of 458,012 galaxies whose minor axis were larger than 14 pixels (semi-minor axis of 7 pixels),
which we subjectively determined was the smallest sized disk on which spiral structure could be observed.
For each galaxy, the number of votes for each of the 6 categories was converted
into a fraction (Table \ref{tab:GZ1properties}).
\begin{table}[hbt]
\begin{tabular}{|l|llllll|l|}
\hline
category     &  EL & EDGE & S-wise & Z-wise & MG & DK & total \\
\hline
Winner by 50\% majority & 261700 & 53873 & 25102 & 23807 & 4431 & 755 & 369668 \\
Percentage & 57.14\% & 11.76\% & 5.48\% & 5.20\% & 0.97\% & 0.16\% & 80.71\% \\
\hline
Winner by max vote & 309591 & 73009 & 33007 & 31340 & 8406 & 2659 & 458012 \\
Percentage & 67.59\% & 15.94\% & 7.21\% & 6.84\% & 1.84\% & 0.58\% & 100\% \\
\hline
\end{tabular}
\caption{The 6 types of votes in Galaxy Zoo 1 across our sample of 458,012 GZ1 galaxies,
along with the fraction of galaxies in each category as voted by the GZ1 humans.
Note that not all galaxies have a winning vote that is a 50\% majority, although every galaxy
has a maximum vote (we ignore ties, which are rare).
}
\label{tab:GZ1properties}
\end{table}

As can be seen in Table \ref{tab:GZ1properties}, there is a significant excess of S-wise
spiral galaxies, using either a majority-vote winner, or a less stringent ``max vote'' winner;
similar surplusses of S-wise spirals are seen using other, more stringent criteria
\citep{Lintott2008,Land2008}.  In our case, using the 50\% majority-wins criterion, there
are $25102+23807=48909$ galaxies with visible spiral structure, but there is an S-wise
excess of $5.86\sigma$ (see Table \ref{tab:final}) compared to 48909 coin flips;
the ``max vote'' criterion shows an even stronger excess, with a statistical significance
of $6.57\sigma$.
As we will see from Table \ref{tab:final} below, the effect gets smaller
as we insist on higher human classification confidence, but never goes away
even when 100\% of humans agree on the chirality of a small set of galaxies.
The statistical significance of this bias is detailed as a function of
human confidence in the first quarter of Table \ref{tab:final}, in which
both the selection of galaxies, and their chirality, are chosen by GZ1 humans.
As can be seen, the bias is detected at a level of somewhere between $3\sigma$
and $6\sigma$, depending upon the human confidence level.

Whether this excess is real or not has been a matter of some debate.
\cite{Lintott2008} and \cite{Land2008} show that the bias seems to disappear if galaxy
images are flipped with 50\% probability before being shown to humans, suggesting that
somehow the humans are biased towards choosing S-wise galaxies.  Whether the bias is
a human cognitive bias, or perhaps due to website design or positioning of the buttons
is unclear, and of little astronomical interest in any case.  However, other studies
\citep{Longo2011,Shamir2012Handedness} have suggested that the bias is real rather than artifactual.

In this paper we put the problem to rest.  We show below that the bias is almost certainly
a human bias, and not a property of the actual GZ1 galaxies.

\section{Nature of the bias}
\label{sec:nature}

\subsection{More S-wise than Z-wise spins for all values of ``spirality''}
\begin{figure}
\includegraphics[width=1\linewidth]{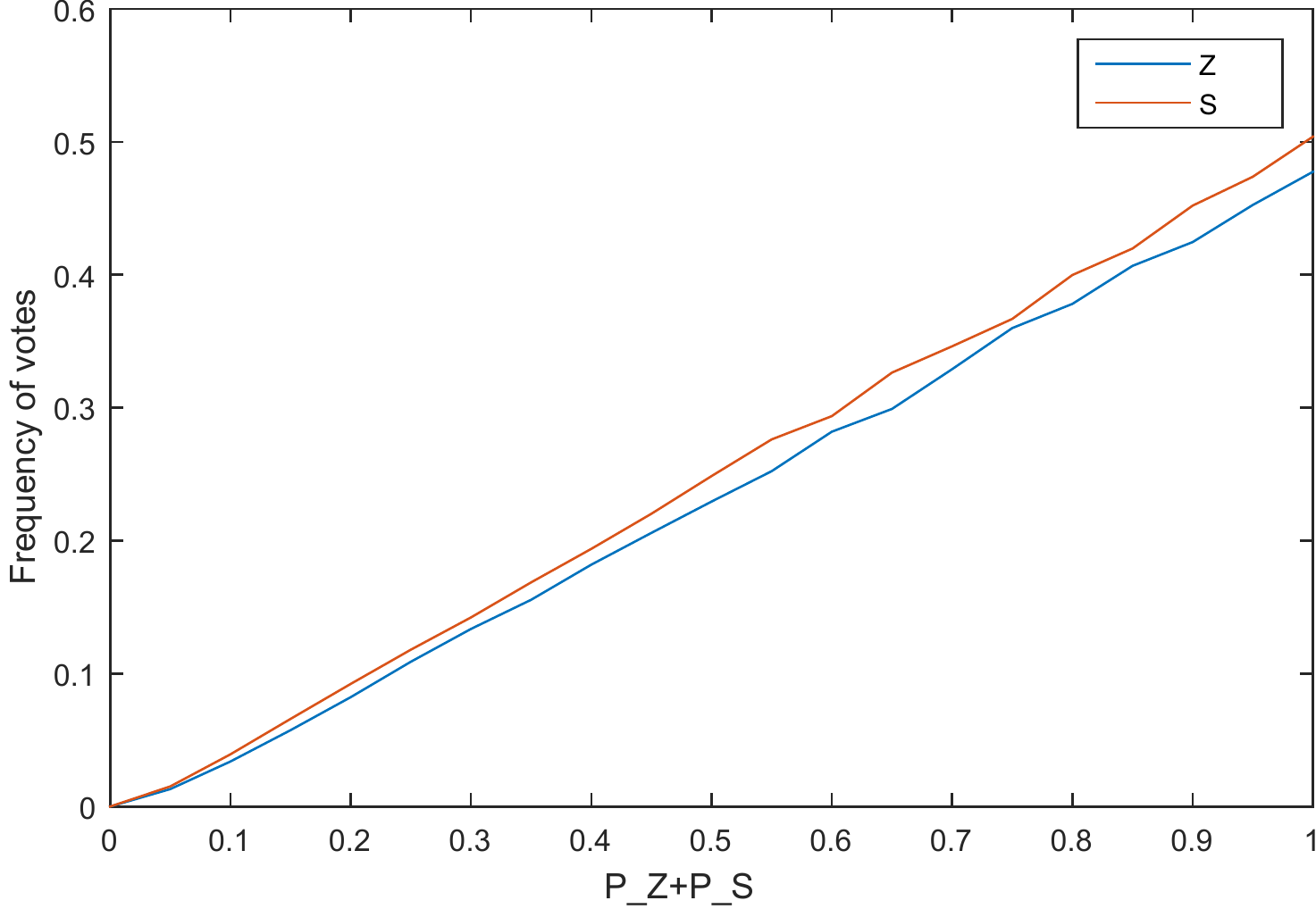}
\caption{Lines joining the frequency histograms (vertical axis) of S-wise and Z-wise
galaxies, according to GZ1 humans, having $x=P_S+P_Z$ (horizontal axis)
among 20 equally-spaced bins in [0,1].
Note that {\em all} galaxies in the entire GZ1
sample are represented in this plot; galaxies to the left end tend to be elliptical or edge-on,
while galaxies to the right end have clearly visible spiral structure.
We see that the S-wise bias manifests across the entire spectrum,
so for example near $x=1$,
we see that among all galaxies for which $P_S+P_Z \ge 0.95$, slightly more than half have $P_S\ge 0.95$,
while slightly less than half have $P_Z\ge 0.95$.
The selection effect manifests because any cutoff in $P_S+P_Z$ that is intended as a threshold above which a galaxy
is considered to have visible spiral structure will automatically include more S-wise than Z-wise galaxies.}
\label{fig:selection}
\end{figure}

Figure \ref{fig:selection} shows the frequencies of galaxies with the two winning
chiralities, as voted by GZ1 humans, as a function of their sum $P_S+P_Z$.
We refer to this sum as the {\em spirality} of a galaxy, and its value
is meant to represent the probability that there exists any observable spiral
structure.\footnote{As distinct from GZ1's $P_{CS}=P_S+P_Z+P_{EDGE}$, which
includes edge-on disks and represents if the galaxy, as seen from any direction, is a disk galaxy.}
As can be seen, the S-wise bias manifests across all
values of spirality even down close to zero, where the galaxies are
unlikely to be spiral at all.  Furthermore, we note that if one
chooses any cutoff in spirality (or similarly $\max(P_S,P_Z)$) meant
to isolate galaxies with visible spiral structure, then any such
sample will automatically include more S-wise than Z-wise galaxies,
because the S-wise curve is uniformly above the Z-wise one for all
values of spirality.
We shall demonstrate, as did
\cite{Land2008} and \cite{Lintott2008}, that this bias is spurious and not reprentative
of the true chirality distribution.

\subsection{Do humans actually disagree on chirality?}

\cite{Land2008} briefly mentioned that there did not
appear to be significant disagreement between humans about chirality.
This statement seems at odds with Figure \ref{fig:selection}.
Here we study that statement in detail, because understanding it may
prove crucial to understanding where the bias comes from.
To test the hypothesis that humans can disagree on the chirality of
a galaxy, we introduce
the idea of the {\em opposing vote}, which we define for any galaxy
as {\it the most popular vote other than the most popular chirality}.
Note that this is not quite the same as the second most popular vote,
because if the most popular {\em chirality} is not the most popular
vote {\em overall}, then the opposing vote is actually the winning vote
overall.  In other words,
\begin{itemize}
\item[1)] When one of the two chiralities is the winning vote,
then the opposing vote is the second most popular vote.
\item[2)] When neither of the two chiralities is the winning vote,
the ``opposing vote'' is the winning vote for that galaxy.
\end{itemize}

\begin{figure}
\begin{center}
\vskip -1cm 
\includegraphics[width=0.7\linewidth]{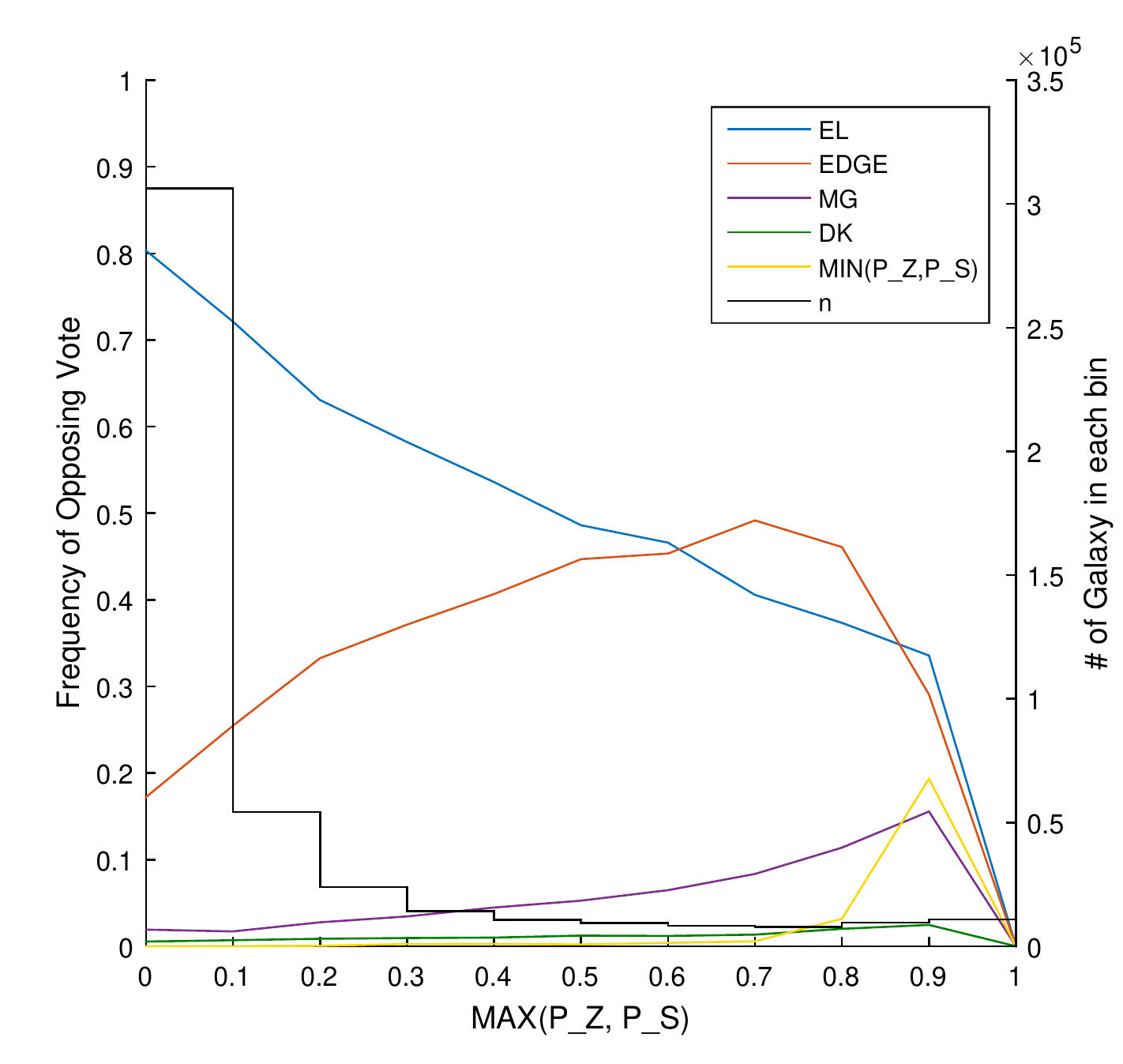}
\includegraphics[width=0.7\linewidth]{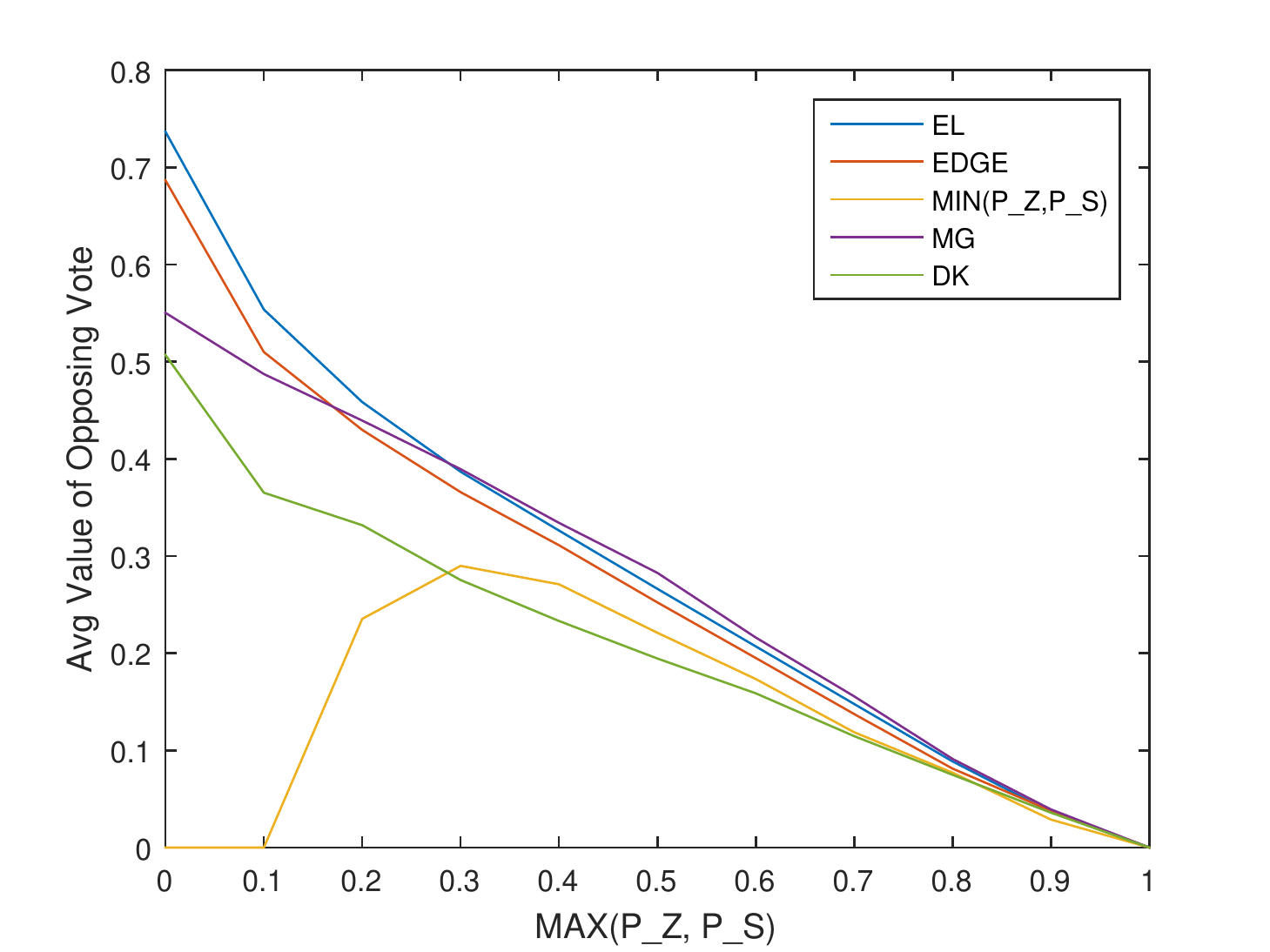}
\caption{{\bf Top}: Histogram of galaxy count (black bars) and frequency of the opposing vote (colored curves),
as a function of the most popular chirality vote.
As can be seen, the losing chirality is rarely the opposing vote.
{\bf Bottom}: Average value of the opposing vote, with the same horizontal axis.}
\label{fig:opposing}
\end{center}
\end{figure}
Figure \ref{fig:opposing} describes the distribution and structure of
the opposing votes, as a function of the most popular chirality, which
is just $\max(P_S,P_Z)$, even if that winning chirality is not the
winning vote across all 6 votes.  The top half of Figure \ref{fig:opposing}
shows the frequency that each of the other 5 votes (which are the losing chirality
plus EL, EDGE, MG, DK) occur as the opposing vote.  The first observation
is that the losing chirality, $\min(P_S,P_Z)$,
is almost never the opposing vote, even for very small values of the
winning vote.  That is to say, humans virtually never disagree
on the chirality of a galaxy; even when only a small percentage of
people actually choose a chirality, they still agree on that chirality.

Instead, the top half of Figure \ref{fig:opposing} demonstrates that
the opposing vote is almost always either EDGE or EL.  This tells
us that the selection effect in Figure \ref{fig:selection} occurs
when people are uncertain whether they see spiral structure at all;
the galaxy may be an edge-on disk galaxy with indistinct spiral
structure, or appear to be elliptical that has faint spiral structure,
but those that choose a chirality in that case tend, for whatever
reason, to be slightly more inclined to choose S-wise over
Z-wise, but even in those cases the humans tend to agree with each
other on the chirality chosen.  In other words, to arrive at the
S-wise bias, humans are ``stealing'' votes from EDGE and EL, not
from Z-wise.  This allows the bias to exist even though humans
virtually never disagree on chirality.
Another interesting observation
of the top half of Figure \ref{fig:opposing} is that near the origin,
the EL and EDGE curves correctly show that, among galaxies that have
no visible spiral structure, about 80\% are elliptical and just
under 20\% are edge-on---in rough agreement with Table \ref{tab:GZ1properties}.

The bottom half of Figure \ref{fig:opposing} shows the average
{\em value} of the opposing vote---i.e., the fraction of people,
per-galaxy, who cast the opposing vote.  As expected, as the winning
chirality approaches 1, the average value of the opposing vote approaches
zero.  Also of interest is the fact that the sum of the winning chirality and
the opposing vote value tends, on average, to be above 70\%, so that
the top two votes take the lion's share of the votes.   We do see,
however, that even though the losing chirality is rarely the
opposing vote, it tends to be a strong second when it does occur;
these are probably galaxies that have strong spiral structure but
are somehow disrupted so as to make the chirality unclear; one may
hazard a guess that they may in fact be advanced mergers.

Finally, again referring to Figure \ref{fig:opposing}, the sharp peak of
the losing chirality (yellow line) at $x=0.9$ in the top figure is not a
problem because, as the lower figure shows, the {\em value} of that vote
is tiny, as is the value of all the other non-winning votes (as they
must be, since the winning chirality is taking 90\% of the votes).

These graphs show that humnans do not significantly
disagree with each other when determining chirality, which is
an observation that is not at all obvious from any of the studies that
have occurred to date.  In fact it would be surprising if humans disagreed
to any significant extent on winding direction, because in all but a very
small number of cases, our own intuitive observation is that if there is
a winding direction at all, it should be fairly obvious.  These graphs
strongly confirm this intuition.  Together, Figures \ref{fig:selection}
and \ref{fig:opposing} demonstrate that the bias has nothing to do with
humans disagreeing on winding direction.  Instead, what is happening is
that whenever there is uncertainty about whether or not there {\em exists}
spiral structure {\em of any chirality}---that is, when a significant
proportion of humans vote either edge-on or elliptical---then those
that {\em do} vote for a chirality tend to vote for S-wise.  The reason
for this is still unknown, but the three obvious choices are
(a) a human visual cortex bias;
(b) something about the design of the web page for the GZ1 survey
induces people to preferentially choose the S-wise button;
or (c) there is a real chirality bias in the SDSS sample of galaxies.

\section{Unbiased machine determination of winding direction}

At this point we have two relevant observations: (1) galaxies voted S-wise
significantly outnumber galaxies voted Z-wise, and (2) humans do not
significantly disagree on chirality. In the absense of evidence for
a human bias, this would directly imply that there is a real chirality
bias in the universe.
\cite{Land2008} convincingly demonstrated that this is not the case
by having the GZ1 humans re-classify a subset of spiral
galaxies while randomly left-right flipping each image with 50\%
probability.  However, beyond demonstrating that the bias was human,
they did not attempt to correct for it on a galaxy-by-galaxy basis.

\begin{figure*}[ht]
\plotsix{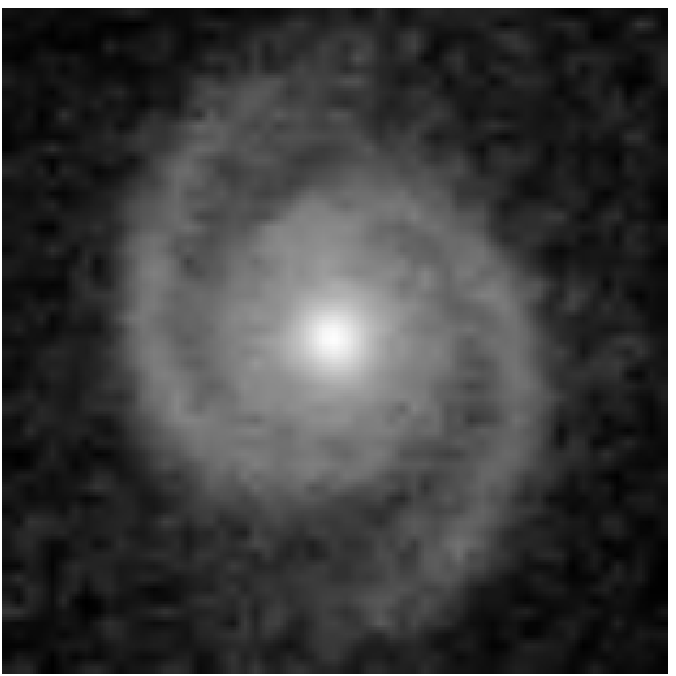}{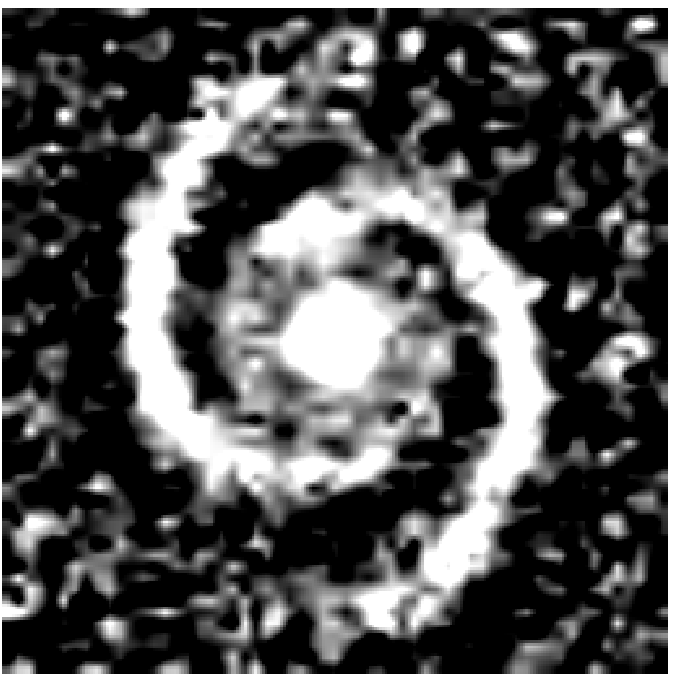}{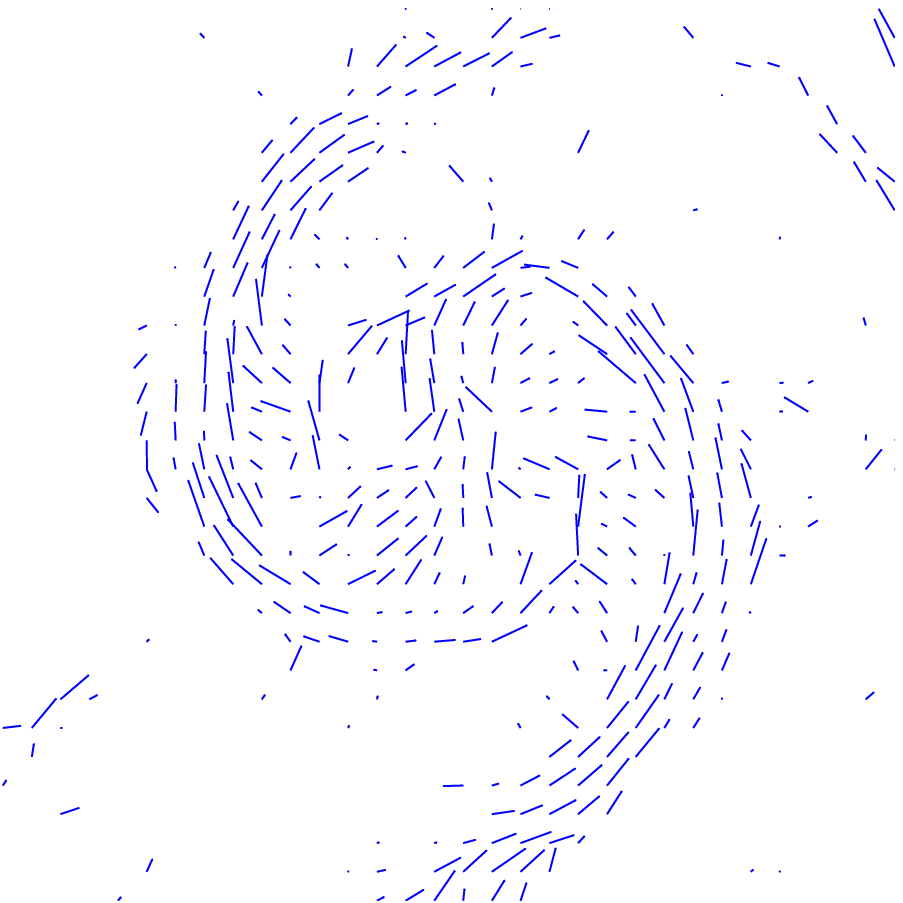}{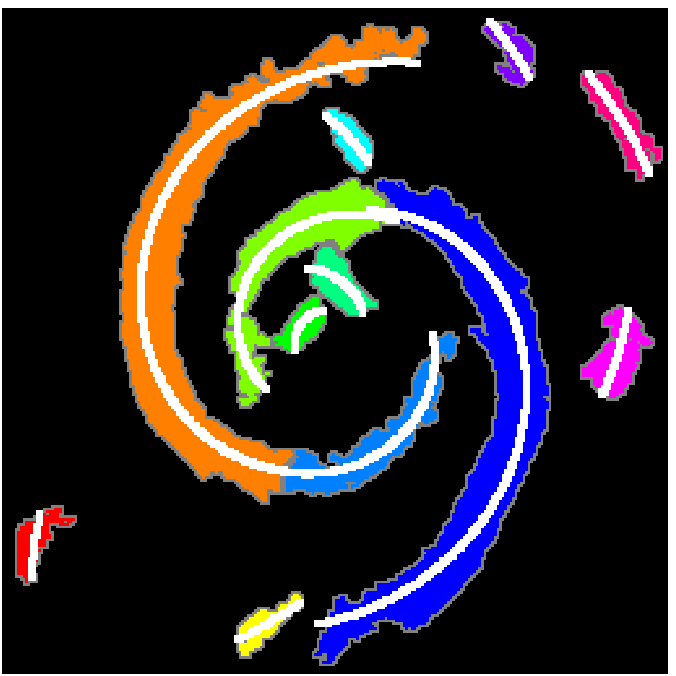}{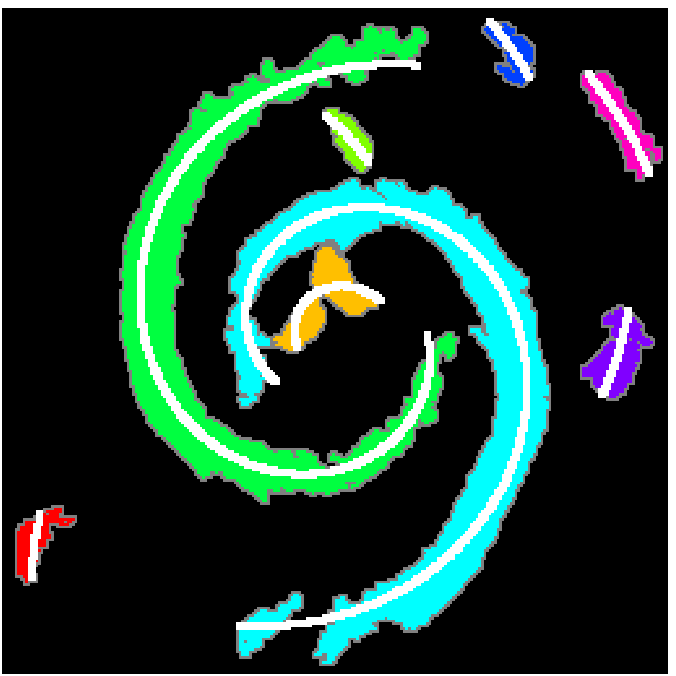}{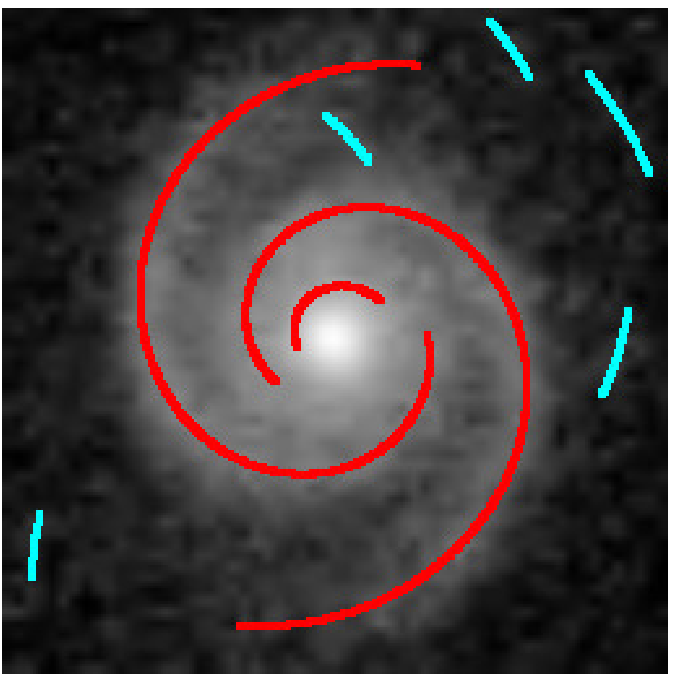}
$\begin{array}{cccccc}
\;\;\;\;\;\;\;\,  ({\bf a}) \;\;\;\;\;\;\;
&\;\;\;\;\;\;\, ({\bf b}) \;\;\;\;\;\;\;\;\;\;\;
&\;\;\;\;\;\; ({\bf c}) \;\;\;\;\;\;\;\;\;\;\;
&\;\;\;\;\;\;\; ({\bf d}) \;\;\;\;\;\;\;\;\;\;
&\;\;\;\;\; ({\bf e}) \;\;\;\;\;\;\;\;\;\;
&\;\;\;\;\;\; ({\bf f}) \;\;\;\;\;\;\;\;\;\;\;
\end{array}$
\caption{Steps SpArcFiRe \citep{DavisHayes2014} takes in describing a spiral galaxy image.
{\bf a)} The centered and de-projected image.
{\bf b)} Contrast-enhanced image.
{\bf c)} Orientation field (at reduced resolution for display purposes).
{\bf d)} Initial arm segments found via Hierarchical Agglomerative Clustering (HAC) of nearby pixels with similar orientations and consistent logarithmic spiral shape, overlaid with the associated logarithmic spiral arcs fitted to these clusters.
{\bf e)} Final pixel clusters (and associated arcs) found by merging compatible arcs.
{\bf f)} Final arcs superimposed on image (a).  Red arcs wind S-wise, cyan arcs wind Z-wise.
}
\label{fig:parse}
\end{figure*}

SpArcFiRe\footnote{{\it SP}iral {\it ARC} {\it FI}nder and {\it RE}porter}
\cite{DavisHayes2014} is an automated method that decomposes
a spiral galaxy into its constituent arms.  A very brief summary of
how SpArcFiRe works is depicted in Figure \ref{fig:parse}.
As described in \cite{DarrenDavisThesis2014,DavisHayes2014},
it was tested on a sample of 29,250 GZ galaxy images chosen by the leader
of the Galaxy Zoo Project\footnote{Stephen Bamford, Personal Communication.
The selection criteria were:
$(GZ1_{P_S} + GZ1_{P_Z}) > 0.8$ OR $(GZ2_{Features Or Disk} > 0.7$ AND $GZ2_{Not Edge On} > 0.7$ AND $GZ2_{spiral} > 0.8)$.}.
Among many other things, one
of SpArcFiRe's outputs is a determination of the galaxy's winding direction.
Given a list of found spiral arcs in a galaxy image, there are many ways
to determine a global winding direction for the entire galaxy.  Some
arcs are longer than others, some may wind in the opposite direction to
the majority, and some ``arcs'' are actually just noise mistaken for
an arc.  We found that the most reliable measure of the winding direction
of the galaxy was a length-weighted vote of the winding direction of
all the discovered spiral arcs.

\begin{figure}
\includegraphics[width=0.7\linewidth]{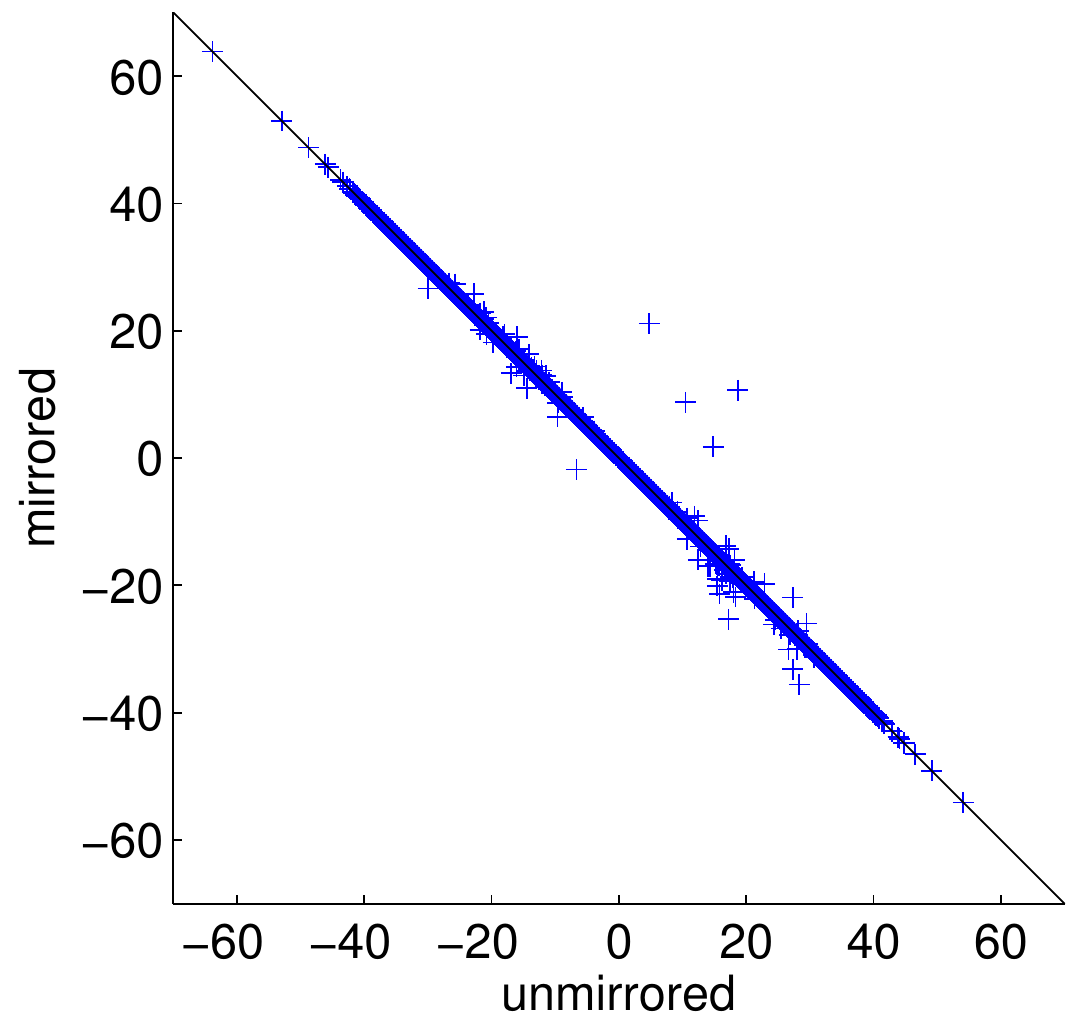}
\caption{Galaxy-level pitch angles reported by SpArcFiRe using unmirrored and left-to-right
mirrored input images across 29,250 ``clear'' spiral galaxies (see text for definition). These
galaxy-level pitch angles are calculated as the arc-length-weighted average of all arcs agreeing
with the dominant winding direction (as determined by an arc-length-weighted vote).
We see that for almost all galaxies the measured pitch angle almost exactly negative, as it should be.
The diagonal line gives $y = -x$; cases on this line are visually underrepresented due to overlap.
More importantly, only 5 cases out of 29,250 disagree on chirality, showing that SpArcFiRe
is chirality-unbiased to a level of almost 1 part in 10,000.}
\label{fig:SFflipped}
\end{figure}

To demonstrate that this measure is unbiased with regard to spin direction,
we refer to Figure \ref{fig:SFflipped}, which shows a scatter plot of the left-to-right
mirrored vs. unmirrored value of the galaxy's pitch angle as measured by SpArcFiRe;
in a perfect reversal, the two should be negatives of each other.
We find that in 29,094 out of 29,250 cases
(99.47\% of cases), the two are negatives of each other
to within $10^{-4}$ degrees.  Even more
relevant to this paper, we find that in all but 5 cases (99.983\% of cases),
the chirality of the mirrored image is correctly flipped compared to the
unmirrored case.  Thus, the chirality determination of SpArcFiRe is
unbiased, with respect to flipped images, to better than 2 parts it $10^4$.

We then ran SpArcFiRe on the entire Galaxy Zoo sample of galaxies, in order
to determine the chirality of galaxies in an unbiased manner.  However, we
still used the human GZ1 determination of $P_{S}+P_{Z}$ to select which
galaxies actually display spiral structure.
The second quarter of Table \ref{tab:final} details the statistical significance
of the winding direction bias, as a function of the {\em human} confidence
in observed spiral structure, but when the chirality is determined by
the unbiased SpArcFiRe algorithm.  The statistical signifance of the
S-wise bias is weaker than in the first quarter of Table \ref{tab:final},
but surprisingly, the bias is still significant to somewhere between
$2\sigma$ and $3\sigma$.

\section{Unbiased machine determination of spirality}
\label{sec:Pedro}

Our goal in this section is to explain how we created a machine learning
algorithm that was capable of reproducing the spirality $P_S+P_Z$, while
being simultaneously unable to reproduce either $P_S$ nor $P_Z$ alone.
That is, we want to create a spirality measure for a galaxy that is
provably independent of chirality.

\subsection{Building a selector that is unbiased to chirality}

As alluded to earlier, the problem is not in the actual determination
of chiraltiy.  Humans do not disagree with each other on chirality, and
in fact the human determination of chirality agrees with the SpArcFiRe
determination of chirality in between 95\% and 98\% of cases on the
GZ1 clean sample, depending upon SpArcFiRe's own determination of its
certainty\citep[tables 5.1 and 5.2, column ``80'']{DarrenDavisThesis2014}, and the
cases of chirality disagreement between GZ1 humans and SpArcFiRe appear
randomly distributed.

Figure \ref{fig:selection} points to the problem: S-wise galaxies
outnumber Z-wise ones for {\em any} set of galaxies selected using
a criterion of either $\max(P_S,P_Z)$ or $P_S+P_Z$ greater than some
threshold $\alpha$, and $P_S$ and $P_Z$ are taken from the human GZ1 vote
values.  Thus, we must determine some method of determining if there is a
selection bias and if so, try to eliminate it.

To do this, we need to create a sample of galaxies that have visible
spiral structure (``spirality''), but selected in a way that is unbiased
to winding direction. To do this we create a machine learning algorithm
that is provided with attributes of the galaxy that are independent of
winding direction, and tell it to attempt to reproduce $P_S+P_Z$. We
then demonstrate that it can reproduce $P_S+P_Z$ with reasonable accuracy
and then show that it is unable to simultaneously reproduce winding direction
to any level better than chance.

In order to create such an algorithm, we need to ensure that the features
it uses (that is, the measurements of the galaxy) are features that are
independent of chirality.  This may not be trivial, as recent work has
suggested that even photometric data may be able to recover winding diretion
to a significant degree \citep{Shamir2016PhotometricAsymmetry}.
We choose our attributes to
include some photometric attributes that were disjoint with those that
\cite{Shamir2016PhotometricAsymmetry} found to be correlated with chirality, in addition to
several SpArcFiRe outputs with all chirality information removed.

Our list of input attributes to our machine learning algorithm,
assumed to be independent of chirality, are as follows. From the SDSS database, we allow
parameters used by \cite{Banerji2010} (colors, de Vaucouleurs fit axial ratios, exponential
fit axial ratios, exponential disk fit log likelihood, de Vaucouleurs fit log likelihood,
star log likelihood, ratios of Petrosian radii, Adaptive shape measures, adaptive
ellipticities, adaptive 4th moment, and a texture parameter), as well as
absolute magnitudes and disk-to-bulge ratios. From SpArcFiRe \citep{DavisHayes2014}
we allow all numerical output parameters including pitch angles after having taken their
absolute value.  Such parameters include counts and lengths of spiral arcs,
the absolute value of their pitch angles, and
the number and length of arcs of agreeing and disagreeing chiralities (with the actual
chirality replaced by "majority" and "minority").

Finally, since it is known that machine learning algorithms tend to reproduce
the input distribution of target values we trained our machine
on a set of galaxies that were 50-50 S-wise and Z-wise according to the
GZ1 humans.

We applied two filters to the data to build a dataset of high-confidence spiral galaxies. For S-wise the rule was $((P_S+P_Z > 0.6)\cap (P_S > 0.5)\cap (P_S - P_Z > 0.3))$; which means that there are at least 60\% of the votes for spirality, at least 50\% of the final votes were for $P_S$, and there is at least 30\% more votes for $P_S$ than for $P_Z$. The first and second rules are effective in filtering out other types of objects and the third in making sure that the humans have a higher agreement not only in spirality but also in chirality. Similarly to build our Z-wise set the rule was $((P_S+P_Z > 0.6)\cap (P_Z > 0.5)\cap (P_Z - P_S > 0.3))$. After this we sampled 19500 objects from each class, to assure we had a balanced dataset, totaling 39000 objects.

We built a random forest model to predict chirality. 
As it is common with these models, we performed what is called a ``hyperparameter'' search across possible machine configurations to decide what was the optimal number of trees per forest and the optimal number of features per tree. We built 45 models with number of trees varying on the interval \{10,15,20,25,30,35,40,45,50\} and number of features per tree from \{30,40,50,60,70\}. We used a Bernoulli distribution to sample 75\% of the data for training the forests and the remaining 25\% to test the models.

\begin{figure}
\includegraphics[width=1\linewidth]{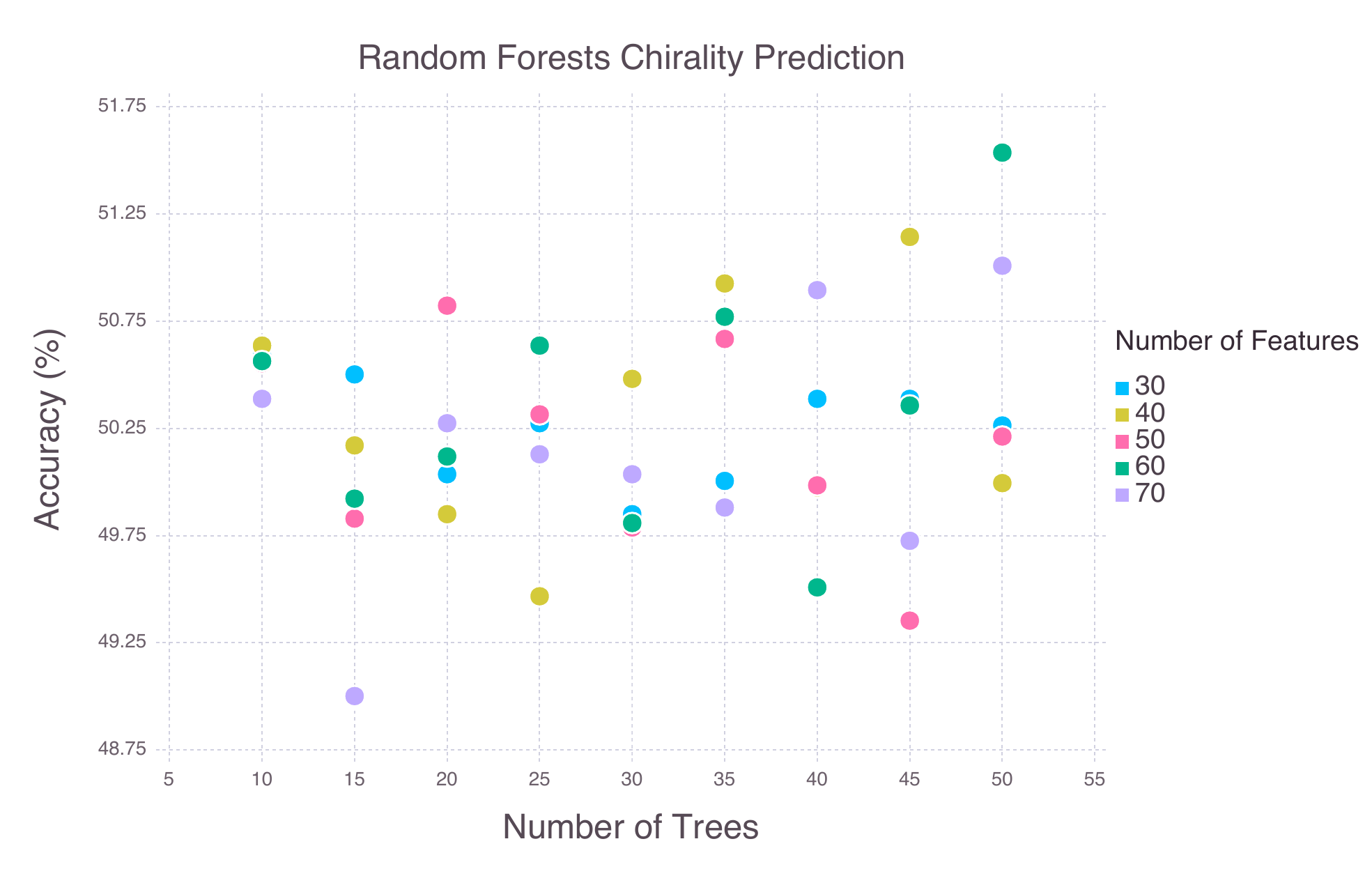}
\caption{Chirality Prediction using Random Forests with 45 different architectures, based on the number of trees and the number of features for each forest.}
\label{fig:ChiPred}
\end{figure}

As mentioned above we expect not to be able to predict chirality with any level of confidence. The accuracy of the models range from from 49\% to 51.25\%, heavily centered around 50, as it is portrayed in Figure \ref{fig:ChiPred}. Due to chance in data sampling and the way that Random Forests are built we expected this variability to occur. Notice that 15, or 1/3 of all the models built, have an accuracy below 50\%, i.e they are worse than a coin flip for tracking chirality. To make sure that our models are not able to indeed predict chirality we decided to further investigate the 3 models that had at least 51\% accuracy.

\begin{figure}
\includegraphics[width=1\linewidth]{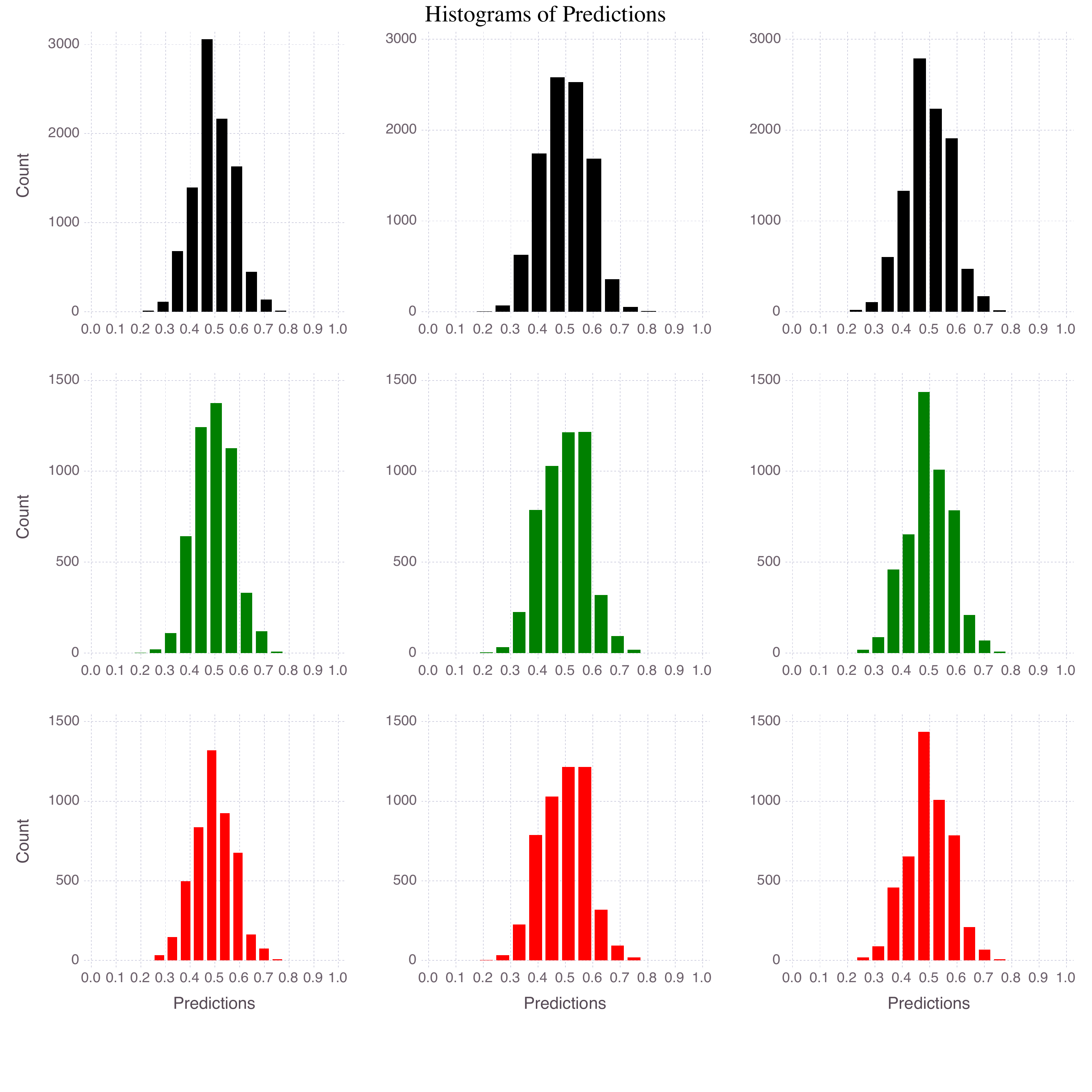}
\caption[Caption for HIS]{Histograms of the predictions for the 3 models that had the highest accuracy. Each column represents a model with the first being the best overall, using 50 trees and 60 features and an accuracy of 51.54\%, the second, using 45 trees and 40 features and an accuracy of 51.14\%, and the third using 50 trees and 70 features and an accuracy of 51.01\%. The first row amounts for all the values predicted by the model, and the second and third rows shows the values that the model predicted correctly and incorrectly, respectively.\protect\footnotemark}
\label{fig:HistPred}
\end{figure}
\footnotetext{This means that the model predicted a value p on the interval $0 < p < 0.5$ when the expected value was 0 and and  $0.5 <= p <= 1$ when the expected value was 1.}

When performing a classification task in machine learning one can have a notion of how confident a model is. For our case we set up the target to be either 0, to predict $P_Z$, or 1, to predict $P_S$. The model outputs a number P within this range: if $P < 0.5$ we say the predicted class for that object was $P_Z$, if $P > 0.5$ the predicted class is $P_S$. The more confident a model is that an object belongs to either class the closer the output value will be to those limits (0 and 1). That means that when a model is not very confident of its output, the values predicted will fall mostly in the middle value of the range. To better visualize this distribution Figure \ref{fig:HistPred} shows histograms from the 3 models that had at least 51\% accuracy. As we had supposed, the distributions are heavily centered, indicating that the model has a very low level of confidence on its predictions. Also, from the 3 models at least 77\% of the predictions were between 0.4 and 0.6, i.e., the model had less than 20\% of confidence on the output for at least 77\% of the objects.

To ensure once and for all that these models performance was due to chance we rebuilt those three models using the same data,the same Bernoulli distribution and the same split for test and training and at the end we got different accuracy values for the 3 models that were previously at least 51\%. The best out of the 3 now has an accuracy of 50.23\%.

We conclude that with these models, given these attributes we are unable to retain any information that correlates in any way to chirality of spiral galaxies.

\subsection{Using the same machine to predict spirality}

Now that we have a list of attributes and a machine that is unable to
predict chirality in the form of either $P_S$ or $P_Z$ alone, we use a
machine with the same input attributes and hyperparameters to reproduce
the sum $P_S+P_Z$, which we term the {\it spirality} of a galaxy.

Given a list of human spirality votes $P_S+P_Z$ for each galaxy, we train
the machine on 75\% of the galaxies and test it on the remaining 25\%; we
do this four times, for four non-overlapping 25\% subsets.  The concatenation
of these four 25\% test subsets constitute our database of machine-determined
spiralities that are independent of chirality.  Given a particular galaxy,
the difference between the human value $P_S+P_Z$ and our predicted spirality
$P_{SP}$ is the error for that galaxy.  The root mean squared error across all
galaxies is a typical measure used to assess the accuracy of a predicted model.
In our case, we were able to produce a machine with an RMSE of 0.137 across
our sample of 450,012 galaxies.  This is quite a bit larger than what
other machines have done; for example the Kaggle winner was able to produce
an RMSE of just 0.07 \citep{GZ2Kaggle2015}. However,
they made no effort to remove human bias, and thus it is not surprising that
they are able to reproduce exactly how the humans voted better than we can.

\section{Results} \label{sec:results}

\begin{table}[hbt]
\center
\begin{tabular}{|l|l|l|rr|lr|}
\hline
spirality & chirality     & spirality&                 &              & sigma &           \\
selector  & determination & cutoff   & $|$S-wise$|$    & $|$Z-wise$|$ & value & $p$-value \\
\hline
GZ1   & GZ1  & 0.4 & \bf 32016 & 30619 & 5.58$\sigma$ & $10^{-8}$ \\
humans&humans& 0.5 & \bf 25625 & 24572 & 4.70$\sigma$ & $10^{-6}$ \\
            && 0.6 & \bf 20952 & 20093 & 4.24$\sigma$ & $10^{-5}$ \\
            && 0.7 & \bf 16631 & 16004 & 3.47$\sigma$ & 0.0002 \\
            && 0.8 & \bf 12444 & 11932 & 3.28$\sigma$ & 0.0004 \\
            && 0.9 & \bf 7774 & 7435 & 2.75$\sigma$ & 0.0030 \\
\hline
GZ1 &SpArcFiRe& 0.4 & \bf 31633 & 31002 & 2.52$\sigma$ & 0.006 \\
humans      && 0.5 & \bf 25417 & 24780 & 2.84$\sigma$ & 0.002 \\
            && 0.6 & \bf 20774 & 20271 & 2.48$\sigma$ & 0.007 \\
            && 0.7 & \bf 16533 & 16102 & 2.39$\sigma$ & 0.010\\
            && 0.8 & \bf 12339 & 12037 & 1.93$\sigma$ & 0.030 \\
            && 0.9 &  \bf 7708 &  7501 & 1.68$\sigma$ & 0.050\\
\hline
\hline
unbiased &GZ1 & 0.4 & 29979 & \bf 30184 & 0.84$\sigma$ & 0.250\\
machine     &humans& 0.5 & \bf 19829 & 19743 & 0.43$\sigma$ & 0.130 \\
            && 0.6 & \bf 13130 & 13093 & 0.23$\sigma$ & 0.400 \\
            && 0.7 & \bf 8510 &  8371 & 1.07$\sigma$ & 0.150\\
            && 0.8 & \bf 5028 &  4895 & 1.34$\sigma$ & 0.100 \\
            && 0.9 & \bf 2231 &  2119 & 1.70$\sigma$ & 0.040\\
\hline
unbiased &SpArcFiRe & 0.4 & \bf 30103 & 30060 & 0.18$\sigma$ & 0.40 \\
machine            && 0.5 & \bf 19800 & 19772 & 0.14$\sigma$ & 0.45 \\
            && 0.6 & 13063 & \bf 13160 & 0.60$\sigma$ & 0.30 \\
            && 0.7 & 8371 & \bf 8510 & 1.07$\sigma$ & 0.15 \\
            && 0.8 & 4895 & \bf 5028 & 1.34$\sigma$ & 0.09 \\
            && 0.9 & 2121 & \bf 2229 & 1.64$\sigma$ & 0.05 \\
\hline
\end{tabular}
\caption{Comparing the statistical significance of the chirality bias.
{\bf Selector}: who selects the sample (GZ1 humans or an unbiased machine learning algorithm);
{\bf chirality determination}: who performs the chirality determination (GZ1 humans or unbiased SpArcFiRe algorithm);
{\bf spirality cutoff}: include only galaxies for which $P_{SP}=P_S+P_Z >$ cutoff;
{\bf S-wise and Z-wise}: number of S-wise and Z-wise galaxies in above defined sample; the over-represented chirality is highlighted in bold;
{\bf sigma and $p$-value}: standard deviation and $p$-value of difference between S- and Z-wise count compared to same number of coin flips.
}
\label{tab:final}
\end{table}

The bottom half of Table \ref{tab:final} shows the results of our
chirality bias study when our unbiased machine (\S \ref{sec:Pedro})
selects galaxies based on predicted spirality. As can be seen,
using this machine to perform selection virtually eliminates the
chirality bias, {\em even if humans still choose the chirality}.
This confirms our statement earlier that the GZ1 humans have a
{\em selection} bias, not a chirality bias.  In fact there is
no significant difference between the two subtables in the lower
half of Table \ref{tab:final}: as long as our machine learning
algorithm performs the selection based on unbiased spirality,
it doesn't matter if the winding direction is determined by humans,
or by SpArcFiRe.  In either case, the S-wise bias is either vastly
reduced, or reversed, apparently at random.

\section{Discussion}

Ideally we would like to integrate our new catalog into the GZ1 catalog so as to publish
a ``corrected'' vote catalog in which the chirality bias has been removed.  However, this
is not as simple as rescaling the $P_S$ and $P_Z$ values to our values. Recall that the
S-wise votes are ``stolen'' from the edge-on and elliptical categories.  Thus, we would
need to re-scale {\em all} the vote values on a galaxy-by-galaxy basis, not just the 
two chirality votes, in order to fully correct the bias.  Furthermore, we would like to
do this in a way that only minimally changes the values of the human votes.  Creating
a machine algorithm that simultaneously removes the bias, and also minimizes the change
in human vote values, is non-trivial, and left for future work.

\section*{Acknowledgements}
\noindent PS was supported by CAPES (Coordination for the Improvement of Higher Education Personnel - Brazil) through the Science Without Borders fellowship for PhD Studies. We thank Leon Cao for an early version of the classifier, and Tina Yang for plotting the Figures describing the GZ1 human handedness bias.

\bibliographystyle{chicago}
\bibliography{ms}

\end{document}

%% file: new-noteset.tex
\newcommand{\Itemize}{\begin{itemize}\setlength{\itemsep}{0pt}}

